\documentclass{mn2e}
\usepackage{times}
\input{psfig.sty}
\newif\ifAMStwofonts

\title{Upper Limits on Central Black Hole Masses of Globular Clusters from Radio Emission and a Possible Black Hole Detection in the Ursa Minor Dwarf Galaxy}

\author[Maccarone, Fender \& Tzioumis] {Thomas J. Maccarone \& Robert
P. Fender\\ Astronomical Institute ``Anton Pannekoek'', University of
Amsterdam, Kruislaan 403, 1098 SJ, Amsterdam, The Netherlands
\newauthor Anastasios K. Tzioumis\\Australia Telescope National Facility, CSIRO, Post Office Box 76, Epping NSW 1710, Australia\\}

\date{}

\begin{document}

\maketitle

\label{firstpage}

\def\simlt{\mathrel{\rlap{\lower 3pt\hbox{$\sim$}}
        \raise 2.0pt\hbox{$<$}}}
\def\simgt{\mathrel{\rlap{\lower 3pt\hbox{$\sim$}}
        \raise 2.0pt\hbox{$>$}}}

\input epsf

\begin{abstract}

Intermediate mass black holes have been alternatively predicted to be
quite common in the centers of globular clusters or nearly impossible
to form and retain in the centers of globular clusters.  As it has
been recently shown that radio observations are currently the most
sensitive observational technique for detecting such objects, we have
obtained new deep radio observations of Omega Cen, and have
re-analyzed older observations of M~15 in hope of constraining the
masses of possible black holes in their centers.  In both cases, upper
limits of about 100 $\mu$Jy are found at GHz frequencies.  We find
that if the Bondi-Hoyle accretion rate truly represents the spherical
accretion rate onto a black hole, then the masses of the black holes
in the centers of these two galaxies are severely constrained - with
mass limits of less than about 100 solar masses in both cases.  If
more realistic assumptions are made based on recent work showing the
Bondi rate to be a severe overestimate, then the data for Omega Cen
are marginally consistent with a black hole of about 1/1000 the
cluster's mass (i.e. about 1000 $M_\odot$).  The data for M~15 then
are only marginally consistent with previous reports of a $\sim2000$
solar mass black hole, and we note that there is considerable hope for
either detecting the black hole or improving this upper limit with
current instrumentation.  Finally, we discuss the possibility that the
radio source near the core of the Ursa Minor dwarf spheroidal galaxy
is a $\sim10^4$ $M_\odot$ black hole.
\end{abstract}


\section{Introduction}

Intermediate mass black holes may represent the link between the
stellar mass black holes seen in the Milky Way (which are probably all
less massive than 20$M_\odot$) and the supermassive black holes
thought to exist in the centers of galaxies (which are mostly more
massive than $10^6 M_\odot$, but see e.g. Greene \& Ho 2004 for a
sample of AGN whose black hole masses are likely to be a few hundred
thousand solar masses and Fillipenko \& Ho 2003 for the discussion of
an especially low mass AGN).  It is possible that the AGN have grown
from $\sim 100-1000 M_\odot$ formed from core collapses of Population
III stars (see e.g. Fryer, Woosley \& Heger 2001).  On the other hand,
another natural place to produce these intermediate mass black holes
is in the centers of globular clusters (Miller \& Hamilton 2002) or
young dense star clusters (Portegies Zwart \& McMillan 2002; Portegies
Zwart et al. 2004; G\"urkan, Freitag \& Rasio 2004).

Several methods have been considered for proving the existence of
these intermediate mass black holes, but to date, there is no
conclusive evidence for the existence of a black hole in the
$10^2$-$10^4$ $M_\odot$ range.  Attempts have been made to associate
some or all of the ultraluminous X-ray sources with intermediate mass
black holes on various bases - X-ray spectra showing cool discs yet
high luminosities compared to what would be expected from a stellar
mass black hole (e.g. Miller et al. 2003), quasi-periodic oscillations
similar to those observed from stellar mass black holes, except at a
frequency an order of magnitude lower (Strohmayer \& Mushotzky 2003),
and the association of one of the brightest ULXs (in fact the same one
showing the low frequency QPO) with a young dense star cluster which
is a plausible birth location for this source (Portegies Zwart et
al. 2004) and a plausible scenario for the evolution of an X-ray
binary with such a black hole has been laid out as well.  

Some work has suggested that gravitational radiation recoil should
lead to the ejection, rather than merger of unequal mass black holes
(e.g. Redmount \& Rees 1989; Favata, Hughes \& Holz 2004; Merritt et
al. 2004; Madau \& Quataert 2004), which would suggest that build-up
of black holes through mergers in the centers of globular clusters
should be ineffective.  The most recent work (Favata et al. 2004)
suggests that the recoil kicks should be of order 10-100 km/sec; since
several of the Milky Way's globular clusters have escape velocities
from their cores as large as 50-100 km/sec (Gnedin et al. 2002), it is
plausible that some clusters can grow black holes through mergers,
although the fraction of such clusters might be rather small (see
e.g. Merritt et al. 2004).  Newtonian dynamical interactions may also
eject black holes by releasing gravitational potential energy when
binaries are hardened (Portegies Zwart \& McMillan 2000).  However,
these calculations are usually simplified, and neglect factors such as
the possible effects of gas, which can be quite important, for
example, in the mergers of supermassive black holes in galaxies
(e.g. Escala et al. 2004).  Given that whether globular clusters can
grow intermediate mass black holes is an important theoretical
question, attempts to search for observational signatures of such
objects are well motivated.

Attempts to identify signatures of the influence of a black hole on
stellar orbits have a rather spotty record.  Claims have been made for
intermediate mass black holes in the globular cluster G1 in M~31 and
in M~15 in our own Galaxy (Gebhardt, Rich \& Ho 2002; van der Marel et
al. 2002; Gerssen et al. 2002), but the statistical significance of
these detections is less than $3\sigma$, and more sophisticated
treatments of stellar dynamics found that the mass concentration
observed at the center of M~15 could be explained by a concentration
of heavy stellar remnants (i.e. neutron stars and white dwarfs -
Baumgardt et al. 2003; Gerssen et al. 2003).  In fact, it has been
argued that many globular clusters may simply not have enough stars to
probe the gravitational potential well enough ever to conclusively
prove the existence of a black hole (Drukier \& Bailyn 2003),
motivating other search techniques.

Long ago it was suggested that black holes in the center of globular
clusters should accrete gas, and perhaps produce observable X-ray
emission (Bahcall \& Ostriker 1975).  Their original suggestion that
IMBHs should dominate the X-ray emission has been proven false by the
detection of orbital periods and Type I X-ray bursts (see Liu, van
Paradijs \& van den Heuvel and references within); however, the basic
idea that an accreting black hole should be detectable from its X-ray
emission has been re-stimulated by the capability of the {\it Chandra
X-ray Observatory} to resolve out sensitively the X-ray emission from
the very center of globular clusters.  Searches have been made for
X-ray emission from the accretion onto possible central black holes,
but with only upper limits found (Grindlay et al. 2001; Ho, Terashima
\& Makishima 2003).  Unfortunately, these upper limits are not
currently particularly constraining of theoretical models, and they
already approach closely the limits of current X-ray instrumentation.

Recent advances in our understanding of accretion and its relation
with relativistic jet formation indicate that the most sensitive
waveband for detecting low luminosity intermediate mass black holes
is, in fact, the radio.  As the luminosity of accretion onto a black
hole decreases, the ratio of radio to X-ray power increases (Gallo,
Fender \& Pooley 2003); the ratio of radio to X-ray power also
increases with increasing black hole mass (Falcke \& Biermann 1996,
1999; Merloni, Heinz \& Di Matteo 2003; Falcke, K\"oerding \& Markoff
2004).  Furthermore, as accretion theory is now suggesting that the
Bondi-Hoyle (1944) rate overestimates the actual accretion rate by 2-3
orders of magnitude (e.g. Perna et al. 2003), the X-ray luminosities
from accretion of the interstellar medium by intermediate mass black
holes in globular clusters are likely to be well below detection
limits of currently existing X-ray observatories, but given the
prediction of Miller \& Hamilton (2002) that the black holes should be
about 0.1\% of the total cluster mass the radio luminosities of the
brightest cluster central black holes may be detectable with existing
instrumentation (Maccarone 2004).  The idea of constraining the masses
of black holes based on combined measurements of their X-ray and radio
emission has also been put forth by Merloni (2004) in the context of
high redshift AGN, where dynamical measurements are also not likely to
be possible for quite some time.  In this paper, we present the upper
limits on the radio luminosity of the nucleus of Omega Centauri from
an Australian Telescope Compact Array (ATCA) observation, and
reconsider the implications of past upper limits on radio emission
from VLA searches for pulsars, and consider the implications of these
results for the possible masses of central black holes.  We also apply
the same methodology in order to discuss the possibility that an
source found by the NRAO VLA Sky Survey (NVSS) may be an intermediate
mass black hole near the center of the of the Ursa Minor dwarf
spheroidal galaxy.

\section{Predicted luminosities}
The method for predicting the radio and X-ray luminosities of an
accreting black hole in a globular cluster was set forth in Maccarone
(2004).  There it was shown that radio emission provides a much more
sensitive means of detecting low luminosity, intermediate (or higher)
mass black holes, as have been speculated to exist in globular
clusters.  Here we briefly review the assumptions and summarize the
results:

\begin{itemize}
\item Black hole mass of 1/1000 of the globular cluster's stellar mass (Miller\& Hamilton 2002)
\item Gas density of 0.15 H cm$^{-3}$, approximately the value estimated from pulsar dispersion measures in M~15 and 47~Tuc, and expected from stellar mass loss (Freire et al. 2001).
\item Accretion at 0.1-1\% of the Bondi rate, due to disk winds
(e.g. Blandford \& Begelman 1999) and/or convection (e.g. Quataert \&
Gruzinov 2000), calculated from numerical simulations of low
luminosity flows (e.g. Hawley, Balbus \& Stone 2001; Igumenshchev,
Narayan \& Abramowicz 2003), and constrained by observations of low
luminosity AGN in elliptical galaxies and the lack of observations of
isolated neutron stars accreting from the interstellar medium (Perna
et al. 2003 and references within).  The low end of this range is in good agreement with the constraints on the accretion rate in Sagittarius A* derived from millimeter polarization measurements (see e.g. Bower et al. 2003).
\item A radiative efficiency in the X-rays of:
\begin{equation}
\eta=(0.1)\times\left(1+\frac{A^2}{2 L_{tot}}-A\sqrt{\frac{A^2}{4 L_{tot}^2}+\frac{1}{L_{tot}}}\right),
\end{equation}
with $A$ a constant to be fitted from observations (and being larger
when the jet's kinetic power is a larger fraction of the total
accretion power), and $L_{tot}$ the radiative plus kinetic luminosity
of the system in Eddington units, as used by Fender, Gallo \& Jonker
(2003) to explain the $L_X-L_R$ correlation obsered by Gallo, Fender
\& Pooley (2003), with the idea being that enough kinetic power is
pumped into a jet to make the accretion flow radiatively inefficient
(see also, e.g. Malzac, Merloni \& Fabian 2004).  Whether the
radiative inefficiency is due to mass and energy loss into a jet or
also partially due to advection into the black hole (e.g. Ichimaru
1977; Narayan \& Yi 1994) is not clearly established by these
relations, and does not affect the results presented here.  We have
set $A=6\times10^{-3}$ for these calculations, which is based on a
conservative estimate of the jet power.  This relation is used to
convert a calcuated accretion rate (from the assumed fraction of the
Bondi-Hoyle rate) into a calculated X-ray luminosity.
\item The fundamental plane relationship among X-ray luminosity, radio luminosity and  black hole mass of Merloni et al. (2003), parameterized for convenient applications to Galactic globular clusters:
\begin{equation}
F_{5 GHz} = 10 \left(\frac{L_x}{3\times10^{31} {\rm ergs/sec}}\right)^{0.6} 
\left(\frac{M_{BH}}{100 M_\odot}\right)^{0.78} \left(\frac{d}{10 {\rm kpc}}\right)^{-2} {\rm {\mu}Jy}.
\label{merloni}
\end{equation}
We note that the relation found by Falcke et al. (2004), which
considered only flat spectrum, low luminosity radio sources like those
we expect to see in the centers of dwarf galaxies or globular
clusters, is consistent with this relation within the uncertainties.
\end{itemize}
We then take the numerical data for the globular cluster's masses and
distances from the Harris catalogue (Harris 1996), compute the X-ray
luminosity under the assumptions given above for the gas density,
fraction of the Bondi-Hoyle accretion rate, and radiative efficiency,
and convert the X-ray luminosity into the radio luminosity using the
fundamental plane relation.  The typical flux values predicted are
less than 40 $\mu$Janskys, even with the accretion rate set to
$10^{-2}$ of the Bondi rate, with the exception of that for Omega Cen,
where the cluster is quite large and reasonably close (see the
tabulation of results in Maccarone 2004).  

\section{Observational Data}

Omega Cen was observed with ATCA simultaneously at 4.8 and 8.6 GHz
over 12hr on 2004 May 8. The array was in the '6C' configuration, with
baselines up to 6km. Data reduction was performed with the MIRIAD
software package (Sault, Teuben \& Wright 1995). Flux and bandpass
calibration was performed on PKS 1934-638, and the phase calibrator
was 1315-46 (SUMSS J131829-462034). No source was detected (limits
given below).

We also consider the upper limits on the flux from a central black
hole in M~15 based on the deep VLA observations of Kulkarni et
al. (1990) and Johnston et al. (1991).  These observations were made
at 1.4 GHz and reached a noise level of 43 $\mu$Jy.  Finally, we also
consider the NVSS detection of a 7.1 mJy radio source within the
$1\sigma$ error box of the center of the Ursa Minor dwarf spheroidal
galaxy.

\section{Comparison of Observations with predictions}

\subsection{Omega Cen}

The ATCA data showed no sources within 10 arcseconds of the center of
Omega Cen.  In naturally weighted maps, the rms noise was 57 $\mu$Jy
at 8.6 GHz, and 38 $\mu$Jy at 4.8 GHz.  We assumed a flat (i.e. $f_\nu
\sim$ const) spectrum (as is likely for a low luminosity accreting
black hole) in order to combine the data sets and reduce the noise.
After doing so, the noise level becomes 32 $\mu$Jy, giving a 3$\sigma$
upper limit of just under 100 $\mu$Jy.  For what seems to be the most
likely set of parameters - a gas density which is the same as that in
M~15 and 47~Tuc, an accretion rate of 0.1\% of the Bondi rate, and a
black hole mass of 0.1\% of the cluster's stellar mass - the
prediction of Maccarone (2004), using the methodology outlined
in section 2, was that the radio flux should be about 0.15 mJy.  That
no detection is made while the predicted level value is 4.5$\sigma$
should not as yet be taken as evidence against the hypothesis that
IMBHs can form in globular clusters.  There are several factors of
$\sim$ a few that could be contributing scatter in the theoretical
predictions - most notably the gas density and temperature, but also
the scatter in the fundamental plane relations of MHDM and FKM, and
possible scatter in the black hole mass-globular cluster mass
relation.  We can say with reasonable confidence that there cannot be
both a black hole with a mass of 1/1000 of the cluster mass {\it and}
accretion at $\sim10^{-2}$ of the Bondi rate (considered to be the
likely upper limit of the accretion rate in the work of Perna et
al. 2003), as the predicted level of radio emission in that case would
be more than 20 times the observed level.  Still, in order to make a
strong statement about whether Omega Cen contains an IMBH, we would
need either a detection, or an upper limit about 10 times lower than
the current upper limit.  Since Omega Cen is far south of the
declinations which can be reached with the VLA, this does not seem
likely unless and until the Square Kilometer Array is built in the
Southern Hemisphere.

\subsection{M~15}
Given that Omega Cen is by far the best candidate in the Southern
Hemisphere for detecting an intermediate mass black hole in a globular
cluster, and that it seems unlikely that it will provide useful limits
without new instrumentation, we turn our attention to the Northern
Hemisphere.  Since M~15 had been claimed to show evidence for a
$\sim2000$ $M_\odot$ black hole, and has been the subject of deep
radio observations to search for pulsars, we consider it as well.  The
3$\sigma$ upper limit on the flux from the center of M~15 is 130
$\mu$Jy (Johnston et al. 1991).  Since the nearest ``bright'' source,
AC 211, is about 2.2 arcseconds away, and the data was taken in VLA A
configuration, with 1.4 arcsecond resolution, confusion is not likely
to be a major problem here, as can be verified from looking at the
image of M~15 presented in Johnston et al. (1991).  Furthermore, M~15
is a core collapse globular cluster, so its central position is quite
well defined.  We again assume that the putative IMBH should have a
flat spectrum, and that hence its radio flux should be below about 130
$\mu$Jy at 5 GHz as well as at 1.4 GHz.  This flux level is already
well above that predicted in Maccarone (2004), even for accretion at
$10^{-2}$ of the Bondi rate.  On the other hand, the predicted radio
luminosity in the model of Maccarone (2004) scales roughly as
$M^{3.2}$, and since the claims exist for a black hole much larger
than 1/1000 of the cluster mass (i.e. a 2500 $M_\odot$ black hole has
been claimed, while 1/1000 the cluster mass is only about 440
$M_\odot$), we can come close to making a useful prediction about the
black hole mass with these data.  These data rule out a black hole
more massive than about 600 $M_\odot$, provided that the systematic
factors discussed above are small.

Since the gas density is well measured in M~15 from pulsars, one
source of uncertainty can be eliminated.  Given that a large fraction
of the scatter in the MHDM relation probably comes from the inclusion
of some steep spectrum radio sources (which any BH accreting from the
ISM of a globular cluster is very unlikely to be), and that the more
restrictive sample of FKM shows somewhat less scatter despite imposing
a theoretical mass-X-ray-radio relation rather than fitting to the
data, it seems likely that even this scatter is not too large.  We are
cautious about interpreting the current upper limits as evidence
against a $\sim2000$ $M_\odot$ black hole in M~15, but we note that
this upper limit, unlike the upper limit on the radio luminosity of
Omega Cen, can easily be reduced by looking at higher frequency (where
the VLA is more sensitive), for longer integrations.  A factor of 5
improvement of the upper limit could be obtained in a quite reasonable
4 hour integration, or in a few years with a much shorter integration
from the Expanded VLA (EVLA).  The project to expand the MERLIN array
should also result in the capability to reach similar flux limits.

\subsection{Ursa Minor dwarf}
The Ursa Minor dwarf spheroidal galaxy is one of the closest dwarf
galaxies to the Earth, at a distance of about 66 kpc.  It is one of
the largest of the Milky Way's dwarf spheroidal galaxies in terms of
mass, with an estimated mass of $2.3\times10^{7} M_\odot$ (see Mateo
1998 and references within for the properties of the Milky Way's dwarf
galaxies, including all the data on dwarf spheroidal galaxies' masses
and distances used in this paper).  The NVSS survey (Condon et
al. 1998) indicates the presence of a 7.1 mJy source 20 arcseconds
from the center of the Ursa Minor dwarf.  While this may seem rather
far from the galaxy's center, we note that this is only 0.02 core
radii for this galaxy, and only 1 $\sigma$ from the measured central
position of the galaxy, using the method of Webbink (1985) to estimate
the uncertainties.  We note that this is an inherent uncertainty in
the galaxy's center when measured from stellar light due to the finite
number of bright stars.

Recent work has suggested that the dwarf spheroidal galaxies might
actually have black holes substantially larger than 1/1000 of their
total mass, if they are the remnants of Population III stars and they
began accreting rapidly soon after forming, and contributed to
re-ionizing the universe (Ricotti \& Ostriker 2004).  Given this
suggestion, we have repeated the calculations of Maccarone (2004), for
black holes of 0.1\%, 1\% and 10\% of the galaxy's mass.  We retain
the assumption made for globular clusters that the gas density should
be about 0.15 H cm$^{-3}$.  This is likely to be an overestimate, as
the dwarf spheroidal galaxies are less dense than globular clusters
and their gas should be more spread out, but we note that at the
present, the constraints on gas densities of dwarf spheroidal galaxies
are rather poor (see e.g. Gallagher et al. 2003).  As we find that for
a fraction of the Bondi rate of $10^{-2}$ the radio flux levels are
all substantially higher than could possibly be allowed by the
observations, we make calculations for only $10^{-3}$ of the Bondi
rate.  We find that the three predicted radio fluxes are 1.7 mJy, 510
mJy, and 97 Jy, respectively.  The predicted X-ray luminosities are
$7\times10^{34}$ ergs/sec, $5\times10^{37}$ ergs/sec, and
$1.5\times10^{40}$ ergs/sec.  We note that {\it Einstein} observed the
Ursa Minor dwarf for about 5.5 kiloseconds, and detected no sources
with positions coincident with the NVSS source.  The upper limit of
those observations was $\sim$$10^{35}$ ergs/sec, if at the distance of
the Ursa Minor galaxy (Markert \& Donahue 1985; see also ROSAT upper
limits at a similar flux level from Zang \& Meurs 2001).  These data
are thus all in accord with a black hole of about 40000 $M_\odot$; if
the gas density is $\sim30$ times lower than in the globular clusters,
then a black hole mass about 4-5 times as large can be accomodated.

\subsubsection{The other dwarf spheroidal galaxies}
To determine the false source probability, we have also looked at the
NVSS catalog around the center of the Ursa Minor galaxy with a search
radius of 20 arcminutes.  In this region, 20 sources are found,
leaving a spurious source probability of about 5\% under the
assumption that any source within 1 arcminute of the center of the
galaxy (the 3$\sigma$ error circle) would be a candidate central IMBH.
We argue that there are few hidden trials here by noting that we would
not expect to detect most of the other Milky Way dwarf spheroidals;
scaling the flux from the Ursa Minor dwarf galaxy, the only other
dwarf spheroidal galaxies massive enough and near enough and far
enough North to be detected in the NVSS are those in Draco,
Sagittarius, Sextans and Fornax, and none of these are detected.  The
other four Milky Way dwarf spheroidal galaxies are those in Sculptor
and Carina (which are too far south to have been observed by the VLA);
and Leo I and Leo II, which are at distances of 205 and 250 kpc,
respectively, and have masses of about half the mass and the same mass
of the Ursa Minor, respectively, leading to fluxes expected to be
about 100 times and about 15 times lower than Ursa Minor's, assuming
that $L$$\propto$$M^{3.2}$, as explained above.

Draco's flux would be expected to be a factor of about 2 lower than
that of the Ursa Minor dwarf because of its greater distance (82 vs 66
kpc), while the Sextans galaxy is further than Draco (86 kpc) and less
massive (by about 20\%), their expected fluxes are thus pushed into
the region where the NVSS suffers from serious incompleteness effects.
The expected flux of Sagittarius would be quite hard to determine; if
the mass is assumed just to be the mass in M~54, the globular
cluster-like object thought to be the core of the Sagittarius dwarf
galaxy, then the calculation of Maccarone (2004) is valid and the
expected flux should be only a few microJanskys.  If, on the other
hand, one scales from the mass of the dwarf, then the fact that this
galaxy is in the process of being tidally destroyed makes estimating
its mass quite difficult.  Furthermore, the gas in the Sagittarius
galaxy is likely to have been stripped during its ongoing encounter
with the Milky Way.  The Fornax dwarf is more problematic; assuming
that galaxy mass and distance are the only important parameters, this
galaxy should actually have a brighter radio source than does Ursa
Minor by a factor of about 3 (it is about 3 times as massive and about
twice as far away).  No detection was made in the NVSS within about 2
arcminutes of the galaxy's center, but a factor of 3 in flux is well
within the uncertainties of our calculations.  Furthermore, we note
that the hidden trial problem is less severe than 3 (to account for
the reasonable chance of detecting a radio source in the centers of
Ursa Minor, Draco \& Fornax) times 5\% (the probability of detecting a
source by chance in the 3$\sigma$ error circle of the Ursa Minor
galaxy); the Ursa Minor galaxy has an error radius for its center
about ten times larger than those of Fornax and Draco, the two other
best candidates (largely because it has a much lower core stellar
density, so a small number of bright stars can affect its
center-of-light location much more); the false detection probabilities
in those galaxies are close to zero, so searching the other two
galaxies adds almost nothing to the probability of finding a source by
chance. We thus believe that the central IMBH hypothesis is more
likely than the background AGN hypothesis, but only marginally so, and
that the best way to make this identification more secure is by
localizing it well enoough to identify a unique optical counterpart to
the radio source, and then to take a spectrum of that object.

\section{Conclusions}

At present, we have shown merely that stricter observational upper
limits can be placed on the maximum mass of a possible accreting black
hole in globular clusters through measurements made in the radio than
through measurements made in the X-rays.  Under the assumptions which
most strictly constrain the presence of intermediate mass black holes
without violating other observational constraints, we can rule out the
presence of a black hole in Omega Cen larger than about 20 $M_\odot$ -
that is we can rule out the presence of an intermediate mass black
hole.  Under more realistic assumptions, though, we cannot rule out a
black hole of a few hundred solar masses.

One other additional caveat is that that accreted gas might have some
angular momentum and hence might form an accretion disk.  In this
case, gas might pile up at some outer radius, and then undergo
outbursts, much like the soft X-ray transients in the Milky Way.  In
this case, the assumption we have made that the instantaneous
accretion rate equals the time averaged accretion rate would be
violated, and substantially lower X-ray and radio luminosities might
be expected most of the time than those we have predicted.  However,
the interstellar gas would have to cool down and condense in order to
form an accretion disk.  This seems to be plausible in galaxies such
as the Milky Way, where there may exist a thin disk as a remnant from
a past strong accretion episode (Nayakshin 2003), but seems less
likely in a globular cluster where it is thought that any IMBHs would
have grown through mergers rather than accretion.

Upper limits which would truly constrain the theoretical suggestions
of Miller \& Hamilton (2002) are likely to come only with the Square
Kilometer Array, or possibly from extremely deep observations made by
the High Sensitivity Array or the Very Large Array.  In Maccarone
(2004), the 15 best candidates for detecting an intermediate mass
black hole through its radio emission; 13 of these are in the Southern
Hemisphere, including the 6 best candidates.  If the detection of
intermediate mass black holes in globular clusters is considered a key
science goal of SKA, then this provides a strong justification for
locating the array south of the equator.

One of the key causes for systematic uncertainty in these measurements
is the uncertainty in the gas density in any globular cluster or gas
poor galaxy which does not contain a large number of pulsars.  While
it is comforting to note that the two globular clusters which contain
large number of pulsars have gas densities in close agreement with the
predictions of stellar mass loss (Freire et al. 2001), passages
through the plane of the Galaxy can strip the clusters of most or all
of their gas, so this will not necessarily be the case in all stellar
systems.  In the ROSAT era, X-ray images were obtained of two of the
Milky Way's dwarf spheroidal galaxies in order to find background
quasars, primarily in order to establish an extragalactic reference
frame for astrometric studies of these galaxies, but it was also noted
that these objects might be useful as probes of the interstellar
medium in the dwarf galaxies, in much the same way that high redshift
quasars are used to study the Lyman $\alpha$ forest (Tinney, Da Costa
\& Zinnecker 1997), but it seems that this technique as not yet
produced any detections of gas.  If this technique can be taken better
advantage of, then it should be possible to convert upper limits on
radio emission from the cores of the dwarf galaxies directly into mass
upper limits in the same way as we have done for the globular
clusters; at the present time, it would be premature to do so, given
the greater uncertainties in the dwarf galaxies' gas densities.  It
should be noted that one dwarf galaxy, that in Scultptor (i.e. the
closest one) does have a substantial gas content (Bouchard, Carignan
\& Maschenko 2003), with a central density at least $10^{-3}$
cm$^{-2}$ in neutral gas, and higher concentrations off center, as the
gas distribution is asymmetric.  The suggestion above that the gas
density might be about 1/30 of that in the globular clusters is thus
in reasonable agreement with the data for the one galaxy with the best
observational constraints.

We have found suggestive evidence for an intermediate mass black hole
in the Ursa Minor dwarf spheroidal galaxy.  This should be followed up
first in X-rays, to determine whether its ``radio mass,'' to use the
terminology of Merloni et al. (2003) is of order $10^4 M_\odot$ before
a strong claim of a detection is made, and an optical counterpart
should be searched for as well.  Since the gas density is probably a
lower in this galaxy than in a typical globular cluster, the true mass
may be higher than suggested by the observed radio luminosity in the
context of our calculation, and hence closer to the $10^{6} M_\odot$
suggested by Riccotti \& Ostriker (2004).  If an X-ray source in the
$10^{33}-10^{35}$ ergs/sec range is found, the case for an
intermediate mass black hole would be strengthened dramatically, and
X-ray luminosities in the lower end of that range would be most likely
to indicate the highest mass black holes.

\section{Acknowledgments}
We thank Dave Meier for pointing out that SKA could be especially
useful for future work in this area.  We also are grateful to the
referee Heino Falcke for useful suggestions which have helped improve
the clarity and content of this paper, and to Eva Grebel for brief but
useful discussions of the gas content of dwarf spheroidal galaxies.

\label{lastpage}

\begin{thebibliography}{}
\bibitem{}Bahcall, J.N. \& Ostriker, J.P., 1975, Nature, 256, 23
\bibitem{}Bondi, H., Hoyle, F., 1944, MNRAS, 104, 273
\bibitem{}Bower, G.C., Wright, M.C.H., Falcke, H. \& Backer, D.C., 2003, ApJ, 588, 331
\bibitem{}Condon, J.J., et al. 1998, AJ, 115, 1693
\bibitem{}Falcke, H. \& Biermann, P.L., 1996, A\&A, 308, 321
\bibitem{}Falcke, H. \& Biermann, P.L., 1999, A\&A, 342, 49
\bibitem{}Falcke, H., K\"ording, E. \& Markoff, S., 2004, A\&A, 414, 895
\bibitem{}Favata, M., Hughes, S.A. \& Holz, D.E., 2004, ApJL, 607, 5L 
\bibitem{}Fender, R.P., Gallo, E. \& Jonker, P.G., 2003, MNRAS, 343, 99L
\bibitem{} Fillipenko, A.V. \& Ho, L.C., 2003, ApJL, 588, 13
\bibitem{}Freire, P.C., Kramer, M., Lyne, A.G., Camilo, F.,
Manchester, R.N. \& D'Amico, N., 2001,ApJL, 557, 105
\bibitem{} Fryer, C.L., Woosley, S.E., \& Heger, A., 2001, apJ, 550, 372
\bibitem{}Gallagher, J.S., Madsen, G.J., Reynolds, R.J., Grebel, E.K. \& Smecker-Hane, T.A., 2003, ApJ, 588, 326
\bibitem{} Gallo, E., Fender, R.P. \& Pooley, g.G., 2003, MNRAS, 344, 60
\bibitem{}Gerssen, J., van der Marel, R.P., Gebhardt, K., Guhathakutra, P., Peterson, R.C., Pryor, C., 2002, AJ, 124, 3270
\bibitem{}Gerssen, J., van der Marel, R.P., Gebhardt, K., Guhathakutra, P., Peterson, R.C., Pryor, C., 2003, AJ, 125, 376
\bibitem{} Gnedin, O., Zhao, H., Pringle, J.E., Fall, S.M., Livio, M. \& Meylan, G., 2002, ApJL, 568, L23 
\bibitem{} Greene, J.E. \& Ho, L.C., 2004, 
\bibitem{}Grindlay, J.E., Heinke, C., Edmonds, R.D., Murray, S.S., 2001, Science, 292, 2290
\bibitem{}G\"urkan, M.A., Freitag, M., Rasio, F.A., 2004, ApJ,604, 632
\bibitem{}Harris, W.E. 1996, AJ, 112, 1487
\bibitem{}Ho, L.C., Terashima, Y., Okajima, T., 2003, ApJL, 587, 35
\bibitem{} Ichimaru, S., 1977, ApJ, 214, 840
\bibitem{} Johnston, H.M., Kulkarni, S.R., Goss, W.M., 1991, ApJL, 382, 89L
\bibitem{}Kulkarni, S., R., Goss, W.M., Wolsczan, A. \& Middleditch, J., 1990, ApJL, 363, L5
\bibitem{} Maccarone, T.J., 2004, MNRAS,  351, 1049
\bibitem{} Madau, P. \& Quataert, E., 2004, ApJL, 606, 17 
\bibitem{} Markert, T.H. \& Donahue, M.E., 1985, ApJ, 297, 564
\bibitem{} Mateo, M., 1998, ARA\&A, 36, 435
\bibitem{}Merloni, A., 2004, MNRAS, in press
\bibitem{}Merloni, A., Heinz, S. \& Di Matteo, T., 2003, MNRAS, 345, 1057
\bibitem{} Merritt, D., Milosavljevic, M., Favata, M., Hughes, s.A.,
\& Holz, D.E., 2004, ApJL, 607, 9
\bibitem{} Miller, J.M., Fabian, A.C. \& Miller, M.C., 2004, ApJ, 607, 931
\bibitem{}Miller, M.C., Hamilton, D.P., 2002, MNRAS, 330, 232
\bibitem{} Narayan, R. \& Yi, I., 1994, ApJL, 428, 13L
\bibitem{}Nayakshin, S., 2003, ANS, 324, 483
\bibitem{}Perna, R., Narayan, R., Rybicki, G., Stella, L., Treves, A., 2003, ApJ, 594, 936
\bibitem{}Portegies Zwart, S.F. \& McMillan, S.L.W., 2000, ApJL, 528, 17L
\bibitem{}Portegies Zwart, S.F. \& McMillian, S.L.W., 2002, ApJ, 576, 899
\bibitem{} Portegies Zwart, S.F., Baumgardt, H., Hut, P., Makino, J. \& McMillan, S.L.W., 2004, Nature, 428, 724
\bibitem{}Ricotti, M. \& Ostriker, J.P., 2004, MNRAS, 352, 547
\bibitem{}Sault R.J., Teuben P.J., Wright M.C.H., 1995, In: Astronomical Data
Analysis Software and Systems IV, eds. Shaw R., Payne H.E., Hayes J.J.E.,
ASP conf. ser. 77, 433
\bibitem{}Tinney, C.G., Da Costa, G.S., Zinnecker, H., 1997, MNRAS, 285, 111
\bibitem{}Quataert, E. \& Gruzinov, A., 2000, ApJ, 539, 809 
\bibitem{} Strohmayer, T.E. \& Mushotzky, R.F., 2003, ApJL, 586, 61
\bibitem{} Webbink, R.F., 1985, IAU Symp. 113, ``Dynamics of Star Clusters,'' 541 (Kluwer:Dordrecht)
\bibitem{}Zang, Z. \& Meurs, E.J.A., 2001, ApJ, 556, 24

\end{thebibliography}
\end{document}